\documentclass[11pt]{report}

\usepackage{amsmath}
\usepackage{amssymb}
\usepackage{hyperref}
\usepackage{graphicx}

\begin{document}

\begin{titlepage}

\begin{flushright}\vspace*{-70pt}
{ULB-TH/06-29, \ gr-qc/0611129}\vspace{30pt}
\end{flushright}

\begin{center}
\Huge{An introduction to the mechanics of black
holes}\\\vspace{30pt} \Large{Lecture notes prepared for the Second
Modave Summer School in Mathematical Physics\vspace{30pt}}
\end{center}
\begin{figure}[!h]
\begin{center}
\resizebox{0.35\textwidth}{!}{\mbox{\includegraphics{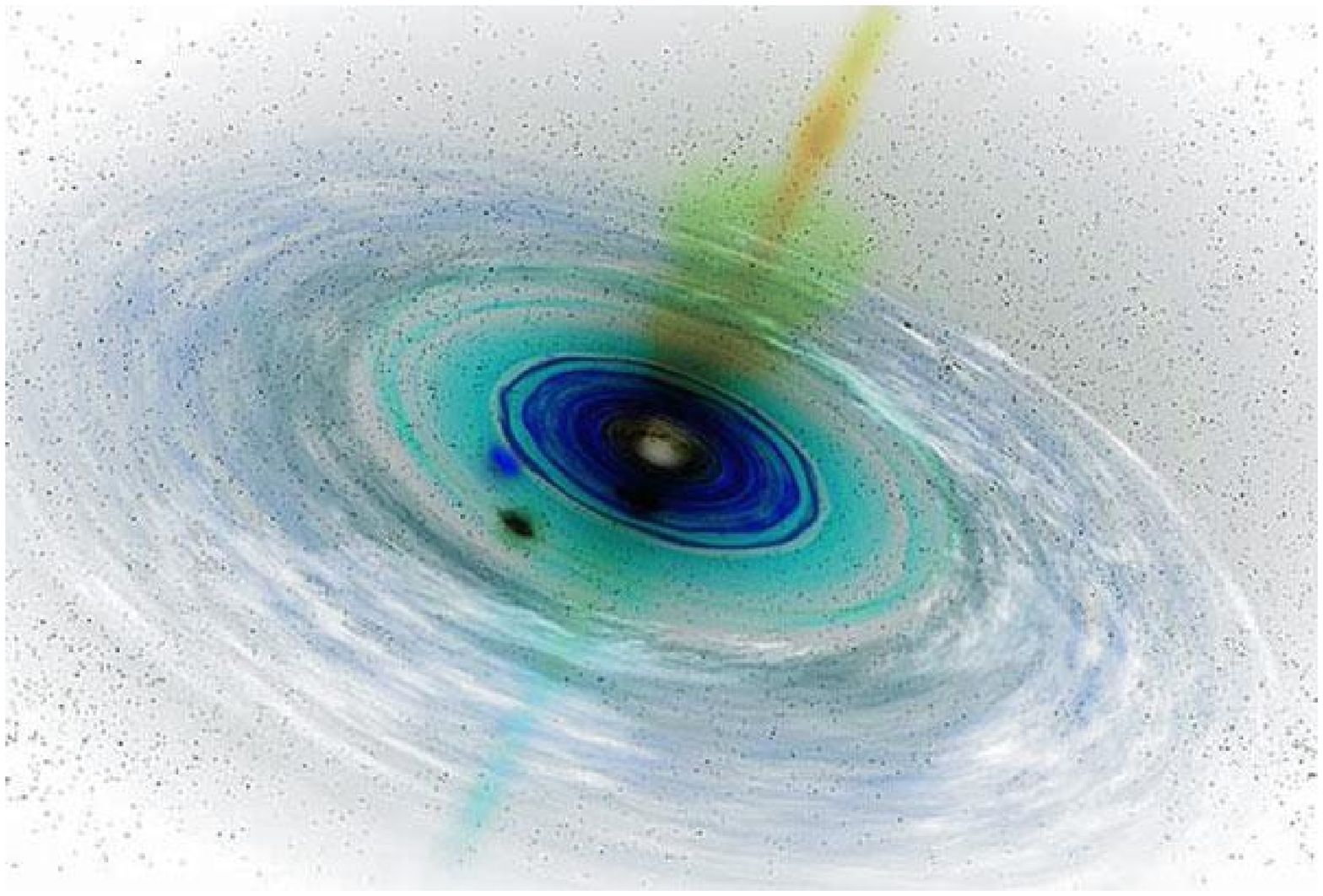}}}
\end{center}
\end{figure}

\begin{center}
Geoffrey Comp\`ere\vspace{10pt}\\
Research Fellow of the National Fund for Scientific
  Research (Belgium)\\
Physique Th\'eorique et Math\'ematique, \\
Universit\'e Libre de Bruxelles and International Solvay Institutes\\
Campus
    Plaine C.P. 231, B-1050 Bruxelles, Belgium\\Email:
    gcompere@ulb.ac.be
\vspace{10pt}
\end{center}

\textbf{Abstract}\hspace{3pt} These notes provide a self-contained
introduction to the derivation of the zero, first and second laws
of black hole mechanics. The prerequisite conservation laws in
gauge and gravity theories are also briefly discussed. An explicit
derivation of the first law in general relativity is performed in
appendix.

\vspace{10pt}\textbf{Pacs:} 04.20.-q, 04.70-s, 11.30.-j

\end{titlepage}

\tableofcontents

\addcontentsline{toc}{chapter}{Introduction}
\chapter*{Introduction}

\emph{Preliminary remark.} These notes (except the third chapter)
are mainly based on previous reviews on thermodynamics of black
holes \cite{Carter1973,Carter:1986ca,Townsend:1997ku,Wald:1999vt}.
\vspace{20pt}

A black hole usually refers to a part of spacetime from which no
future directed timelike or null line can escape to arbitrarily
large distance into the outer asymptotic region. A white hole or
white fountain is the time reversed concept which is believed not
to be physically relevant, and will not be treated.

More precisely, if we denote by $\gimel^+$ the future asymptotic
region of a spacetime $(\mathcal M,g_{\mu\nu})$, e.g. null
infinity for asymptotically flat spacetimes and timelike infinity
for asymptotically anti-de Sitter spacetimes, the black hole
region $\mathcal B$ is defined as
\begin{equation}
\mathcal B \equiv \mathcal M - I^-(\gimel^+),
\end{equation}
where $I^-$ denotes the chronological past. The region
$I^-(\gimel^+)$ is what is usually referred to as the \emph{domain
of outer communication}, it is the set of points for which it is
possible to construct a future directed timelike line to arbitrary
large distance in the outer region.

\begin{figure}[hbtp]
\begin{center}
\resizebox{0.35\textwidth}{!}{\mbox{\includegraphics{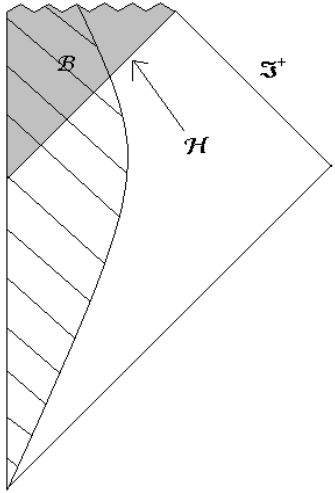}}}
\caption{\small{Penrose diagram of an asymptotically flat
spacetime with spherically symmetric collapsing star. Each point
is a $n-2$-dimensional sphere. Light rays propagate along
$45^\circ$ diagonals. The star region is hatched and the black
hole region is indicated in grey.}} \label{GEO:fig2}
\end{center}
\end{figure}

The \emph{event horizon} $\mathcal{H}$ of a black hole is then the
boundary of $\mathcal B$. Let us denote $J^-(U)$ the causal past
of a set of points $U \subset \mathcal M$ and $\bar J^-(U)$ the
topological closure of $J^-$. We have $I^-(U) \subset J^-(U)$. The
(future) event horizon of $\mathcal M$ can then equivalently be
defined as
\begin{equation}
\mathcal H \equiv \bar J^-(\gimel^+) - J^-(\gimel^+),
\end{equation}
i.e. the boundary of the closure of the causal past of $\gimel^+$.
See Fig.~\ref{GEO:fig2} for an example. The event horizon is a
concept defined with respect to the entire causal structure of
$\mathcal M$.

The event horizons are null hypersurfaces with peculiar
properties. We shall develop their properties in section 1, what
will allow us to sketch the proof of the area theorem
\cite{Hawking:1971vc}. The area theorem also called the second law
of black hole mechanics because of its similarity with classical
thermodynamics \cite{Bekenstein:1972tm} is concerned with the
dynamical evolution of sections of the event horizon at successive
times.

In section 2, we introduce the notion of Killing horizon. This
concept is adapted for black holes in equilibrium in stationary
spacetimes. We will show how the zero law of mechanics, the
consistency of a specific quantity defined on the horizon,
directly comes out of the definitions.

The tools necessary to handle with the conservations laws in
gravity theories are briefly introduced in section 3. In section
4, we derive the first law for two infinitesimally close
equilibrium black holes \cite{Bardeen:1973gs}.

Remark that the zero and first law of black hole mechanics may
also be generalized to black holes in non-stationary spacetimes.
This was done in the framework of ``isolated horizons'' very
recently \cite{Ashtekar:2000sz,Ashtekar:2004cn}. However, in this
introduction, we limit ourselves to the original notion of Killing
horizon.

\vspace{15pt}\noindent\emph{Notation} In what follows,
$\partial_\mu f = f_{,\mu}$ is the partial derivative, while
$D_\mu f = f_{;\mu}$ denotes the covariant derivative.

\chapter{Event horizons}

\section{Null hypersurfaces}

Let $S(x^\mu)$ be a smooth function and consider the $n-1$
dimensional null hypersurface $S(x) = 0$, which we denote by
$\mathcal{H}$. This surface will be the black hole horizon in the
subsequent sections. It is a null hypersurface, i.e. such that its
normal $\xi^\mu \sim g^{\mu\nu}\partial_\nu S$ is null,
\begin{equation}
\xi^\mu \xi_\mu \overset{\mathcal H}{=}0.\label{GEO:xi0}
\end{equation}
The vectors $\eta^\mu$ tangent to $\mathcal{H}$ obey $\eta_\mu
\xi^\mu |_\mathcal{H} = 0$ by definition. Since $\mathcal{H}$ is
null, $\xi^\mu$ itself is a tangent vector, i.e.
\begin{equation}
\xi^\mu = \frac{dx^\mu(t)}{dt}
\end{equation}
for some null curve $x^\mu(t)$ inside $\mathcal{H}$. One can then
prove that $x^\mu(t)$ are null geodesics\footnote{Proof: Let
$\xi_\mu = \tilde f S_{,\mu}$. We have
\begin{eqnarray}
\xi^\nu \xi_{\mu ; \nu} &=& \xi^\nu \partial_\nu \tilde f S_{,\mu}
+
\tilde f \xi^\nu S_{,\mu;\nu}\nonumber \\
&=& \xi^\nu \partial_\nu \ln \tilde f \xi_\mu + \tilde f \xi^\nu
S_{;\nu;\mu}\nonumber  \\
&=& \xi^\nu \partial_\nu \ln \tilde f \xi_\mu +\tilde f \xi^\nu (\tilde f^{-1}\xi_\nu )_{;\mu} \nonumber \\
&=& \xi^\nu \partial_\nu \ln \tilde f \xi_\mu +
\frac{1}{2}(\xi^2)_{,\mu} - \partial_\mu \ln \tilde f \xi^2.
\label{GEO:last1}
\end{eqnarray}
Now, as $\xi$ is null on the horizon, any tangent vector $\eta$ to
$\mathcal{H}$ satisfy $(\xi^2)_{;\mu}\eta^\mu=0$. Therefore,
$(\xi^2)_{;\mu} \sim \xi_\mu$ and the right-hand side of
\eqref{GEO:last1} is proportional to $\xi_\mu$ on the horizon. }
\begin{equation}
\xi^\nu \xi^\mu_{\,\,\, ; \nu} \overset{\mathcal H}{=} \kappa
\xi^\mu,  \label{GEO:geo}
\end{equation}
where $\kappa$ measure the extent to which the parameterization is
not affine.  If we denote by $l$ the normal to $\mathcal{H}$ which
corresponds to an affine parameterization $l^\nu l^\mu_{\,\,\, ;
\nu} = 0$ and $\xi = f(x)\, l$ for some function $f(x)$, then
$\kappa = \xi^\mu \partial_\mu \ln|f|$.

According to the Frobenius' theorem, a vector field $v$ is
hypersurface orthogonal if and only if it satisfies
$v_{[\mu}\partial_\nu v_{\rho]} =0$, see e.g. \cite{Wald:1984rg}.
Therefore, the vector $\xi$ satisfies the irrotationality
condition
\begin{equation}
\xi_{[\mu}\partial_\nu \xi_{\rho]}\overset{\mathcal H}{=}
0.\label{GEO:irr}
\end{equation}

A congruence is a family of curves such that precisely one curve
of the family passes through each point. In particular, any smooth
vector field define a congruence. Indeed, a vector field define at
each point a direction which can be uniquely ``integrated'' along
a curve starting from an arbitrary point.

\begin{figure}[hbtp]
\begin{center}
\resizebox{0.35\textwidth}{!}{\mbox{\includegraphics{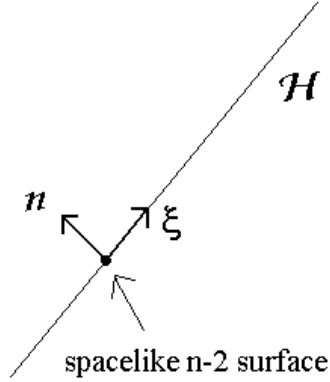}}}
\caption{\small{The null vector $n$ is defined with respect to
$\xi$.}} \label{GEO:fig3}
\end{center}
\end{figure}

Since $S(x)$ is also defined outside $\mathcal H$, the normal
$\xi$ defines a congruence but which is a null congruence only
when restricted to $\mathcal{H}$. In order to study this
congruence outside $\mathcal{H}$, it is useful to define a
transverse null vector $n^\mu$ cutting off the congruence with
\begin{equation}
n^\mu n_\mu = 0, \qquad n_\mu \xi^\mu = -1.\label{GEO:norm}
\end{equation}
The normalization $-1$ is chosen so that if we consider $\xi$ to
be tangent to an outgoing radial null geodesic, then $n$ is
tangent to an ingoing one, see Fig.~\eqref{GEO:fig3}. The
normalization conditions~\eqref{GEO:norm} (imposed everywhere,
$(n^2)_{;\nu} = 0 =(n \cdot \xi)_{;\nu}$) do not fix uniquely $n$.
Let us choose one such $n$ arbitrarily. The extent to which the
family of hypersurfaces $S(x) = const$ are not null is given by
\begin{equation}
\varsigma \equiv \frac{1}{2} (\xi^2)_{;\mu} n^\mu \neq
0.\label{GEO:defvarsig}
\end{equation}

The vectors $\eta$ orthogonal to both $\xi$ and $n$,
\begin{equation}
\eta^\mu\xi_\mu = 0 = \eta^\mu n_\mu,\label{GEO:tg_small}
\end{equation}
span a $n-2$ dimensional spacelike subspace of $\mathcal{H}$. The
metric can be written as
\begin{equation}
g_{\mu\nu} = - \xi_\mu n_\nu - \xi_\nu n_\mu +
\gamma_{\mu\nu}\label{GEO:metric}
\end{equation}
where $\gamma_{\mu\nu} = \gamma_{(\mu\nu)}$ is a positive definite
metric with $\gamma_{\mu\nu}\xi^\mu = 0 = \gamma_{\mu\nu}n^\mu$.
The tensor $\gamma^\mu_{\;\, \nu} =
g^{\mu\alpha}\gamma_{\alpha\nu}$ provides a projector onto the
$n-2$ spacelike tangent space to $\mathcal H$.

For future convenience, we also consider the hypersurface
orthogonal null congruence $l^\mu$ with affine parameter $\tau$
that is proportional to $\xi^\mu$ on $\mathcal H$\footnote{We
shall reserve the notation $\xi^\mu$ for vectors coinciding with
$l^\mu$ on the horizon but which are not null outside the
horizon.},
\begin{eqnarray}
l^\mu l_\mu = 0, \qquad l^\nu l^\mu_{\;\,;\nu}=0, \qquad l^\mu
\overset{\mathcal H}{\sim} \xi^\mu.
\end{eqnarray}
The vector field $l$ extends $\xi$ outside the horizon while
keeping the null property.

\section{The Raychaudhuri equation}

In this section, we shall closely follow the reference
\cite{Carter:1986ca}. We introduce part of the material needed to
prove the area law.

Firstly, let us decompose the tensor $D_\mu \xi_\nu$ into the
tensorial products of $\xi$, $n$ and spacelike vectors $\eta$
tangent to $\mathcal H$ ,\footnote{Proof: Let us first decompose
$D_\mu \xi_\nu$ as
\begin{equation}
D_\mu \xi_\nu = v_{\mu\nu} + n_\mu (C_1 n_\nu + C_2 \xi_\nu + C_3
\eta_{\nu}) + \tilde \eta_\mu \xi_\nu+ \hat \eta_\mu n_\nu -
\xi_\mu \alpha_\nu,
\end{equation}
where $v_{\mu\nu} =
\gamma^\alpha_{\;\,\mu}\gamma^\beta_{\;\,\nu}v_{\alpha\beta}$ and
$\eta^\mu$, $\tilde \eta^\mu$, $\hat \eta^\mu$ are spacelike
tangents to $\mathcal H$. Contracting with $\xi^\mu$ and
using~\eqref{GEO:geo}, we find $C_1=0=C_3$, $C_2 = -\kappa$.
Contracting with $\gamma^\mu_{\;\,\alpha}n^\nu$, we find $\tilde
\eta_\mu = -\gamma^\alpha_{\;\,\mu}n^\beta D_\alpha \xi_\beta$.
Contracting with $\gamma^\mu_{\;\,\alpha}\xi^\nu$, we find finally
$\hat \eta_\mu = -1/2 \gamma^\alpha_{\;\,\mu}D_\alpha(\xi^2) = 0$
thanks to \eqref{GEO:xi0}.}
\begin{equation}
D_\mu \xi_\nu \overset{\mathcal H}{=} v_{\mu\nu} - \xi_\nu (\kappa
n_\mu + \gamma^\alpha_{\;\,\mu}n^\beta D_\alpha \xi_\beta) -
\xi_\mu n^\alpha D_\alpha \xi_\nu,\label{GEO:decomp}
\end{equation}
where the orthogonal projection $v_{\mu\nu} =
\gamma^\alpha_{\;\,\mu}\gamma^\beta_{\;\,\nu}D_\alpha \xi_\beta$
can itself be decomposed in symmetric and antisymmetric parts
\begin{equation}
v_{\mu\nu} = \theta_{\mu\nu} + \omega_{\mu\nu}, \qquad
\theta_{[\mu\nu]}=0,\qquad \omega_{(\mu\nu)}=0.
\end{equation}
The Frobenius irrotationality condition \eqref{GEO:irr} is
equivalent to $\omega_{\mu\nu}|_\mathcal{H} = 0$\footnote{Proof:
We have
\begin{equation}
\xi_{[\mu}\partial_\nu \xi_{\rho ]} = \xi_{[\mu}D_\nu \xi_{\rho]}
= \xi_{[\mu } v_{\nu\rho ]} = \xi_{[\mu} \omega_{\nu\rho]}.
\end{equation}}. The tensor $\theta_{\mu\nu}$ is interpreted as the
expansion rate tensor of the congruence while its trace $\theta =
\theta_{\mu}^{\;\,\mu}$ is the divergence of the congruence. Any
smooth $n-2$ dimensional area element evolves according to
\begin{equation}
\frac{d}{d t}(d\mathcal A) = \theta \,d \mathcal
A.\label{GEO:incr}
\end{equation}
The shear rate is the trace free part of the strain rate tensor,
\begin{equation}
\sigma_{\mu\nu} = \theta_{\mu\nu} - \frac{1}{n-2}\theta
\gamma_{\mu\nu}.
\end{equation}
Defining the scalar $\sigma^2 =
(n-2)\sigma_{\mu\nu}\sigma^{\mu\nu}$, one has
\begin{equation}
\xi_{\mu;\nu}\xi^{\nu;\mu} \overset{\mathcal H}{=}
\frac{1}{n-2}(\theta^2+\sigma^2) + \kappa^2
+\varsigma^2,\label{GEO:eqxixi}
\end{equation}
where $\varsigma$ was defined in \eqref{GEO:defvarsig}. Note also
that the divergence of the vector field has three contributions,
\begin{equation}
\xi^\mu_{\;\,;\mu} \overset{\mathcal H}{=} \theta + \kappa -
\varsigma.\label{GEO:derContr}
\end{equation}
Now, the contraction of the Ricci identity
\begin{equation}
v^\alpha_{\; ;\mu ;\nu} - v^\alpha_{\; ;\nu;\mu} = -R^\alpha_{\;\,
\lambda \mu\nu }v^\lambda,
\end{equation}
implies the following identity
\begin{equation}
(v^\nu_{\;\, ;\nu})_{;\mu}v^\mu = (v^\nu v^\mu_{\; ;\nu})_{;\mu} -
v^{\nu;\mu}v_{\mu;\nu}-R_{\mu\nu}v^\mu v^\nu,\label{GEO:id}
\end{equation}
valid for any vector field $v$. The
formulae~\eqref{GEO:eqxixi}-\eqref{GEO:derContr} have their
equivalent for $l$ as
\begin{eqnarray}
l_{\mu;\nu}l^{\nu;\mu}=
\frac{1}{n-2}(\theta_{(0)}^2+\sigma_{(0)}^2), \qquad
l^\mu_{\;\,;\mu} = \theta_{(0)} ,
\end{eqnarray}
where the right hand side are expressed in terms of the expansion
rate $\theta_{(0)} = \theta \frac{dt}{d\tau}$ and shear rate
$\sigma_{(0)} = \sigma \frac{dt}{d\tau}$ with respect to the
affine parameter $\tau$. The identity \eqref{GEO:id} becomes
\begin{equation}
\frac{d\theta_{(0)}}{d\tau} = \dot\theta_{(0)}  \overset{\mathcal
H}{=} -\frac{1}{n-2}(\theta_{(0)}^2+\sigma_{(0)}^2) -
R_{\mu\nu}l^\mu l^\nu,\label{GEO:Ray}
\end{equation}
where the dot indicate a derivation along the generator. It is the
final form of the Raychaudhuri equation for hypersurface
orthogonal null geodesic congruences in $n$ dimensions.

\section{Properties of event horizons}

As we have already mentioned, the main characteristic of event
horizons is that they are null hypersurfaces. In the early
seventies, Penrose and Hawking further investigated the generic
properties of past boundaries, as event horizons. We shall only
enumerate these properties below and refer the reader to the
references \cite{HawkingEllis,Townsend:1997ku} for explicit
proofs. These properties are crucial in order to prove the area
theorem.

\begin{enumerate}
\item \emph{Achronicity property.} No two points of the horizon can be connected by a timelike
curve.

\item The null geodesic generators of $\mathcal H$
may have past end-points in the sense that the continuation of the
geodesic further into the past is no longer in $\mathcal H$.

\item The generators of $\mathcal H$ have no future end-points, i.e. no
generator may leave the horizon.

\end{enumerate}

The second property hold for example for collapsing stars where
the past continuation of all generators leave the horizon at the
time the horizon was formed. As a consequence of properties 2 and
3, null geodesics may enter $\mathcal H$ but not leave it.

\section{The area theorem}

The area theorem was initially demonstrated by Hawking
\cite{Hawking:1971vc}. We shall follow closely the reviews by
Carter \cite{Carter:1986ca} and Townsend \cite{Townsend:1997ku}.
The theorem reads as follows.
\newpage
\begin{description} \item {\bf Theorem 1} \emph{Area law.}
If
\begin{itemize}
\item[(i)] Einstein's equations hold with a matter stress-tensor
satisfying the null energy condition, $T_{\mu\nu}k^\mu k^\nu \geq
0$, for all null $k^\mu$,
\item[(ii)] The spacetime is ``strongly asymptotically predictable''
\end{itemize}
then the surface area $\mathcal A$ of the event horizon can never
decrease with time.
\end{description}
The theorem was stated originally in $4$ dimensions but it is
actually valid in any dimension $n \geq 3$.

In order to understand the second requirement, let us remind some
definitions. The future domain of dependence $D^+(\Sigma)$ of an
hypersurface $\Sigma$ is the set of points $p$ in the manifold for
which every causal curve through $p$ that has no past end-point
intersects $\Sigma$. The significance of $D^+(\Sigma)$ is that the
behavior of solutions of hyperbolic PDE's \emph{outside}
$D^+(\Sigma)$ is not determined by initial data on $\Sigma$. If no
causal curves have past end-points, then the behavior of solutions
inside $D^+(\Sigma)$ is entirely determined in terms of data on
$\Sigma$. The past domain of dependence $D^-(\Sigma)$ is defined
similarly.

A Cauchy surface is a spacelike hypersurface which every
non-spacelike curve intersects exactly once. It has as domain of
dependence $D^+(\Sigma) \cup D^- (\Sigma)$ the manifold itself. If
an open set $\mathcal N$ admits a Cauchy surface then the Cauchy
problem for any PDE with initial data on $\mathcal N$ is
well-defined. This is also equivalent to say that $\mathcal N$ is
globally hyperbolic.

The requirement (ii) means that it should exist a globally
hyperbolic submanifold of spacetime containing both the exterior
spacetime \emph{and} the horizon. It is equivalent to say there
exists a family of Cauchy hypersurfaces $\Sigma(\tau)$, such that
$\Sigma(\tau^\prime)$ is inside the domain of dependence of
$\Sigma(\tau)$ if $\tau^\prime > \tau$.

Now, the boundary of the black hole is the past event horizon
$\mathcal H$. It is a null hypersurface with generator $l^\mu$
(that is proportional to $\xi$ on $\mathcal H$). We can choose to
parameterize the Cauchy surfaces $\Sigma(\tau)$ using the affine
parameter $\tau$ of the null geodesic generator $l$.

The \emph{area of the horizon} $\mathcal A(\tau)$ is then the area
of the intersection of $\Sigma(\tau)$ with $\mathcal H$. We have
to prove that $\mathcal A(\tau^\prime) > \mathcal A(\tau)$ if
$\tau^\prime
> \tau$.

\vspace{5pt}\noindent {\bf Sketch of the proof:\vspace{5pt}}\\
The Raychaudhuri equation for the null generator $l$ reads
as~\eqref{GEO:Ray}. Therefore, wherever the energy condition
$R_{\mu\nu}l^\mu l^\nu \geq 0 $ hold, the null generator will
evolve subject to the inequality
\begin{equation}
\frac{d\theta_{(0)}}{d \tau} \leq -\frac{1}{n-2}\,\theta_{(0)}^2,
\end{equation}
except on possible singular points as caustics. It follows that if
$\theta_{(0)}$ becomes negative at any point $p$ on the horizon
(i.e. if there is a convergence) then the null generator can
continue in the horizon for at most a finite affine distance
before reaching a point $p$ at which $\theta_{(0)} \rightarrow
-\infty$, i.e. a point of infinite convergence representing a
caustic beyond which the generators intersects.

\begin{figure}[htbp]
\begin{center}
\resizebox{0.35\textwidth}{!}{\mbox{\includegraphics{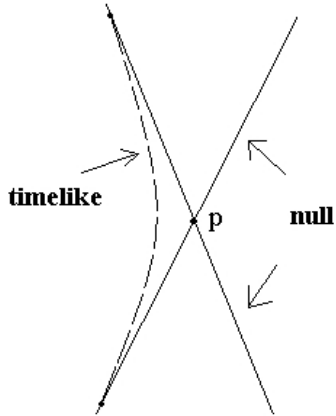}}}
\caption{\small{If two null generators of $\mathcal H$ cross, they
may be deformed to a timelike curve.}} \label{GEO:fig1}
\end{center}
\end{figure}

Now, from the third property of event horizons above, the
generators cannot leave the horizon. Therefore at least two
generators cross at $p$ inside $\mathcal H$ and, following Hawking
and Ellis (Prop 4.5.12 of \cite{HawkingEllis}), they may be
deformed to a timelike curve, see figure~\ref{GEO:fig1}. This is
however impossible because of the achronicity property of event
horizons. Therefore, in order to avoid the contradiction, the
point $p$ cannot exist and $\theta_{(0)}$ cannot be negative.

Since (at points where the horizon is not smooth) new null
generators may begin but old ones cannot end,
equation~\eqref{GEO:incr} implies that the total area $\mathcal
A(\tau)$ cannot decrease with increasing $\tau$,
\begin{equation}
\frac{d}{d\tau}\mathcal A \geq \oint \theta_{(0)} \,d\mathcal A
\geq 0.
\end{equation}
This completes the proof.

In particular, if two black holes with area $\mathcal A_1$ and
$\mathcal A_2$ merge then the area $\mathcal A_3$ of the combined
black hole have to satisfy
\begin{equation}
\mathcal{A}_3 > \mathcal{A}_1+\mathcal{A}_2.
\end{equation}
The area $A(\tau)$ do not change if $\theta = 0$ on the entire
horizon $\mathcal H$. The black hole is then stationary.

Note that this derivation implicitly assume regularity properties
of the horizon (as piecewise $C^2$) which may not be true for
generic black holes. Recently these gaps in the derivation have
been totally filled in \cite{Chrusciel:2000cu,Wald:1999vt}.

\chapter{Equilibrium states}

\section{Killing horizons}

In any stationary and asymptotically flat spacetime with a black
hole, the event horizon is a Killing horizon \cite{HawkingEllis}.
This theorem firstly proven by Hawking is called the rigidity
theorem. It provides an essential link between event horizons and
Killing horizons.\footnote{The theorem further assumes the
geometry is analytic around the horizon. Actually, there exist a
counter-example to the rigidity theorem as stated in Hawking and
Ellis \cite{HawkingEllis} but under additional assumptions such as
global hyperbolicity and simple connectedness of the spacetime,
the result is totally valid \cite{Chrusciel:1996bj}.}

A Killing horizon is a null hypersurface whose normal $\xi$ is a
Killing vector
\begin{equation}
\mathcal{L}_\xi g_{\mu\nu} = \xi_{\mu ; \nu} + \xi_{\nu ; \mu} =
0.\label{GEO:Kill}
\end{equation}
This additional property will allow us to explore many
characteristics of black holes.

The parameter $\kappa$ which we call now the surface gravity of
$\mathcal{H}$ is defined in~\eqref{GEO:geo}. In asymptotically
flat spacetimes, the normalization of $\kappa$ is fixed by
requiring $\xi^2 \rightarrow -1$ at infinity (similarly, we impose
$\xi^2 \rightarrow -\frac{r^2}{l^2}$ in asymptotically anti-de
Sitter spacetimes).

For Killing horizons, the expansion rate $\theta_{\mu\nu} =
\gamma_{(\mu}^{\;\,\alpha}\gamma_{\nu)}^{\;\,\beta}D_{\alpha}\xi_\beta
= 0$, so $\theta = \sigma = 0$. Moreover,
from~\eqref{GEO:derContr} and \eqref{GEO:Kill}, we deduce
$\varsigma = \kappa$. Equation~\eqref{GEO:eqxixi} then provides an
alternative definition for the surface gravity,
\begin{equation}
\kappa^2 = -\frac{1}{2}\xi_{\mu ; \nu} \xi^{\mu;\nu}|_{\mathcal
H}.
\end{equation}
Contracting~\eqref{GEO:geo} with the transverse null vector $n$,
one has also
\begin{equation}
\kappa = \xi_{\mu;\nu}\xi^\mu n^\nu |_\mathcal{H} = \frac{1}{2}
(\xi^2)_{,\mu} n^\mu |_\mathcal{H}.\label{GEO:kappa}
\end{equation}
The Raychaudhuri equation~\eqref{GEO:Ray} also states in this case
that
\begin{equation}
R_{\mu\nu}\xi^\mu \xi^\nu \overset{\mathcal H}{=}
0,\label{GEO:RayEQ}
\end{equation}
because $l$ is proportional to $\xi$ on the horizon.

From the decomposition~\eqref{GEO:decomp}, the irrotationality
condition~\eqref{GEO:irr} and the Killing property
$\xi_{[\mu;\nu]} = \xi_{\mu;\nu}$, one can write
\begin{equation}
\xi_{\mu;\nu} \overset{\mathcal H}{=} \xi_\mu q_\nu - \xi_\nu
q_\mu,\label{GEO:defq}
\end{equation}
where the covector $q_\mu$ can be fixed uniquely by the
normalization $q_\mu n^\mu = 0$. Using \eqref{GEO:kappa}, one can
further decompose the last equation in terms of
$(n,\xi,\{\eta\})$ as
\begin{equation}
\xi_{\mu;\nu} \overset{\mathcal H}{=} -\kappa (\xi_\mu n_\nu -
\xi_\nu n_\mu) + \xi_\mu \hat \eta_\nu - \hat \eta_\mu
\xi_\nu,\label{GEO:ortho}
\end{equation}
where $\hat \eta$ satisfy $\hat \eta \cdot \xi = 0 = \hat \eta
\cdot n$. In particular, it shows that for any spacelike tangent
vectors $\eta$, $\tilde \eta$ to $\mathcal H$, one has
$\xi_{\mu;\nu}\eta^\mu \tilde \eta^\nu \overset{\mathcal H}{=} 0$.

\section{Zero law}

We are now in position to prove that the surface gravity $\kappa$
is constant on the horizon under generic conditions. More
precisely,
\begin{description} \item {\bf Theorem 2} \emph{Zero law.} \cite{Bardeen:1973gs}
If
\begin{itemize}
\item[(i)] The spacetime $(M,g)$ admits a Killing vector $\xi$
which is the generator of a Killing horizon $\mathcal H$,
\item[(ii)] Einstein's equations hold with matter satisfying the
dominant energy condition, i.e. $T_{\mu\nu}l^\nu$ is a
non-spacelike vector for all \mbox{$l^\mu l_\mu \leq 0$},
\end{itemize}
then the surface gravity $\kappa$ of the Killing horizon is
constant over $\mathcal H$.
\end{description}

Using the aforementioned properties of null hypersurfaces and
Killing horizons, together with
\begin{equation}
\xi_{\nu;\mu;\rho} = R_{\mu\nu\rho}^{\,\,\,\,\,\,\,\,\,\,
\tau}\xi_\tau,
\end{equation}
which is valid for Killing vectors, one can obtain (see
\cite{Carter:1986ca} for a proof)
\begin{eqnarray}
\dot \kappa = \kappa_{,\mu}\xi^\mu &\overset{\mathcal H}{=}& 0,\\
\kappa_{,\mu}\eta^\mu &\overset{\mathcal H}{=}& -
R_{\mu\nu}\xi^\mu\eta^\nu,
\end{eqnarray}
for all spacelike tangent vectors $\eta$. Now, from the dominant
energy condition, $R_{\mu\nu}\xi^\mu$ is not spacelike. However,
the Raychaudhuri equation implies \eqref{GEO:RayEQ}. Therefore,
$R_{\mu\nu}\xi^\mu$ must be zero or proportional to $\xi_\nu$ and
$R_{\mu\nu}\xi^\mu\eta^\nu = 0$.

This theorem has an extension when gravity is coupled to
electromagnetism. If the Killing vector field $\xi$ is also a
symmetry of the electromagnetic field up to a gauge
transformation, $\mathcal L_\xi A_\mu +\partial_\mu \epsilon = 0$,
one can also prove that the electric potential
\begin{equation}
\Phi = -(A_\mu \xi^\mu+\epsilon)|_{\mathcal H}
\end{equation}
is constant on the horizon.

\chapter{Conservation laws}

\hspace{20pt}``\textit{Anybody who looks for a magic formula for
``local gravitational energy-momentum'' is looking for the right
answer to the wrong question. Unhappily, enormous time and effort
were devoted in the past to trying to ``answer this question''
before investigators realized the futility of the enterprise}''
\begin{flushright}
Misner, Thorne and Wheeler \cite{Misner:1970aa}
\end{flushright}

According to Misner, Thorne and Wheeler, the Principle of
Equivalence forbids the existence of a localized energy-momentum
stress-tensor for gravity. No experiment can be designed to
measure a notion of local energy of the gravitational field
because in a locally inertial frame the effect of gravity is
locally suppressed. However, it is meaningful to ask what is
energy content of \emph{region} or of the \emph{totality} of a
spacetime. For the first issue, we refer the reader to the
literature on recent quasi-local methods
\cite{Brown:1992bq,Brown:1992br}, see also \cite{Chang:1998wj} for
the link with pseudo-tensors. Here, we shall deal with the second
issue, the best studied and oldest topic, namely, the definition
of global conservation laws for gravity (and for general gauge
theories).

It exists an overabundant literature over conservations laws. Some
methods are prominent but none impose itself as the best one, each
of them having overlapping advantages, drawbacks and scope of
application. The following presentation will therefore reflect
only a biased and narrow view on the topic.

\section{The generalized Noether theorem}

Let us begin the discussion by recalling the first Noether
theorem.
\begin{description} \item {\bf Theorem 3} \emph{First Noether Theorem}
Any equivalence class of continuous global symmetries of a
lagrangian $L\, d^n x$ is in one-to-one correspondence with an
equivalence class of conserved currents $J^\mu$, $\partial_\mu
J^\mu = 0$.
\end{description}
Here, two global symmetries are equivalent if they differ by a
gauge transformation and by a symmetry generated by a parameter
vanishing on-shell. Two currents $J^\mu$, $J^{\prime \mu}$ are
equivalent\footnote{In $n=1$ dimension, two currents differing by
a constant $J \in \mathbb{R}$ are also considered as equivalent.}
if they differ by a trivial current,
\begin{equation}
J^\mu \sim J^{\prime \mu} + \partial_\nu k^{[\mu\nu]} +
t^\mu(\frac{\delta L}{\delta \phi}), \qquad t^\mu \approx 0,
\end{equation}
where $t^\mu$ depends on the equations of motion (i.e. vanishes
on-shell). This theorem is essential in order to define the energy
in classical mechanics or in field theories. For example, the
total energy of the field associated with a time translation
$(\partial_t)^\mu$ on a spacelike Cauchy surface $\Sigma$ is
defined as $E = \int_\Sigma  J_\mu  n^\mu$ where $n^\mu$ is the
unit normal to $\Sigma$ and $J_\mu = T_{\mu\nu} (\partial_t)^\nu$
where $T_{\mu\nu}$ is the conserved stress-tensor of the field
($\partial_\mu T^{\mu\nu} = 0$).

In diffeomorphic invariant theories, the infinitesimal coordinate
transformations generated by a vector $\xi$ are pure gauge
transformations. The first Noether theorem implies that \emph{all
currents $J_\xi$ associated to infinitesimal diffeomorphisms are
trivial}\footnote{For example, in general relativity, one has
$\delta L = \delta g_{\mu\nu} \frac{\delta L}{\delta g_{\mu\nu}} +
\partial_\mu\Theta^\mu(g,\delta g) $ for some $\Theta(g,\delta g)$. For a diffeomorphism, one
has $\delta g_{\mu\nu} = D_\mu \xi_\nu + D_\nu \xi_\mu$, $\mathcal
L_\xi L = \partial_\mu ( \xi^\mu L)$ and $\delta
g_{\mu\nu}\frac{\delta L}{\delta g_{\mu\nu}} =
-2\sqrt{-g}D_{\mu}\xi_\nu  = -2\partial_{\mu}(\sqrt{-g}
G^{\mu\nu}\xi_\nu)$ where $G_{\mu\nu}$ is the Einstein tensor.
What is usually called the Noether current is $J^\mu =
\Theta^\mu(g,\mathcal L_\xi g) - \xi^\mu L$. However, it is
trivial because (as implied by Noether theorem) it exists a
$k^{\mu\nu}= k^{[\mu\nu]}$ such that $J^\mu = -2\sqrt{-g}
G^{\mu\nu}\xi_\nu + \partial_\nu k^{[\mu\nu]}$.}.

The main lesson of Theorem 3 is that for gauge theories, one
should \emph{not} look at conserved currents. In order to
generalize the Noether theorem, it is convenient to introduce the
notation for $n-p$ forms,
\begin{equation}
(d^{n-p}x)_{\mu_1 \dots \mu_{n-p}} = \frac{1}{p!
(n-p)!}\epsilon_{\mu_1 \dots \mu_n} dx^{\mu_{n-p+1}} \wedge \cdots
\wedge dx^{\mu_n}.
\end{equation}
The Noether current can be reexpressed as a $n-1$ form as $J =
J^\mu (d^{n-1}x)_\mu$ which is closed, i.e. $dJ = 0$.

Now, we shall see that the conservation laws for gauge theories
are \emph{lower degree conservation laws}, involving conserved
$n-2$ forms, $k = k^{[\mu\nu]} (d^{n-2}x)_{\mu\nu}$, i.e. such
that $dk = 0$ or $\partial_\nu k^{[\mu\nu]} = 0$. Indeed, in the
nineties, the following theorem was proved
\cite{Barnich:1995db,Barnich:1995ap}, see also
\cite{Bryant:1995,Andersonbook} for related work and
\cite{Torre:1997cd} for introductions to local cohomology,
\begin{description} \item {\bf Theorem 4} \emph{Generalized Noether Theorem}
Any parameter of a gauge transformation vanishing on-shell such
that the parameter itself is non zero on-shell is in one-to-one
correspondence with $n-2$-forms $k$ that are conserved on-shell,
$d k \approx 0$ (up to trivial $n-2$-forms\footnote{Trivial
$n-2$-forms include superpotentials that vanish on-shell and
topological superpotentials that are closed off-shell, see
\cite{Torre:1997cd} for a discussion of topological conservation
laws.} and up to the addition of the divergence of a $n-3$-form).
\end{description}
Essentially, the theorem amounts first to identify the class of
gauge transformations vanishing on-shell with non-trivial gauge
parameters with the cohomology group $H_2^n(\delta | d)$ and the
class of non-trivial conserved $n-2$ forms with $H_0^{n-2}(d |
\delta)$. The proof of the theorem then reduces to find an
isomorphism between these cohomology classes.

As a first example, in electromagnetism (that may be coupled to
gravity), the trivial gauge transformations $\delta A_\mu =
\partial_\mu c = 0$ are generated by constants $c \in \mathbb{R}$.
The associated $n-2$-form is simply
\begin{equation}
k_c[A,g] = c \; (d^{n-2}x)_{\mu\nu} \sqrt{-g} F^{\mu\nu}
\end{equation}
which is indeed closed (outside sources that are not considered
here) when the equations of motion hold. The closeness of $k$
indicate that the electric charge
\begin{equation}
\mathcal Q_E = \oint_S k_{c=1},
\end{equation}
i.e. the integral of $k$ over a closed surface $S$ at constant
time, does not depend on time and may be freely deformed in vacuum
regions\footnote{Explicitly, let $S$ be some surface $r=const$,
$t=const$. The $r$ component of $dk = 0$ is $\partial_t
k^{tr}+\partial_A k^{Ar} = 0$ where $A=\theta,\phi$ are the
angular coordinates. The time derivative of $\mathcal Q$ is then
given by $\oint_S
\partial_t k^{tr} (d^{n-2}x)_{tr} = -\oint_S \partial_A k^{Ar}
d\mathcal A = 0$ by Stokes theorem.
Similarly, the $t$ component of $dk = 0$ is $\partial_r
k^{rt}+\partial_A k^{At} = 0$ and $\partial_r \mathcal Q = 0$
too.}.

In generally covariant theories, the trivial diffeomorphisms
$\delta g_{\mu\nu} = \mathcal L_{\xi}g_{\mu\nu}$ are generated by
the Killing vectors\footnote{Trivial diffeomorphisms must also
satisfy $\mathcal L_\xi \phi^i = 0$ if other fields $\phi^i$ are
present.}
\begin{equation}
\mathcal L_{\xi}g_{\mu\nu} = \xi_{\mu ; \nu} + \xi_{\nu ; \mu} =
0.
\end{equation}

However, here comes the problem: there is \emph{no} solution to
the Killing equation for \emph{arbitrary} fields. Therefore,
\emph{none} vector $\xi$ can be associated to a \emph{generically}
conserved $n-2$ form. The hope to associate a conserved quantity
\begin{equation}
\mathcal Q_\xi[g] = \oint_S K_\xi[g],\qquad d K_\xi \approx 0,
\qquad K_\xi \,\,\backslash \hspace{-10pt}\approx 0,
\end{equation}
for a given vector $\xi$ to all solutions $g$ of general
relativity is definitively annihilated.

What is the way out? In fact, there are many ways out with
different methods and results. In these notes, we will weaken the
requirements enormously by selecting special surfaces $S$, vectors
$\xi$ and classes of metrics $g$ and by allowing the conserved
quantities
\begin{equation}
\mathcal Q_\xi[g,\bar g] = \oint_S K_\xi[g,\bar g],\qquad d K_\xi
\approx 0,\qquad K_\xi \,\,\backslash \hspace{-10pt}\approx
0,\label{charge_gen}
\end{equation}
to depend on some background solution $\bar g$. If we normalize
the charge of the background (typically Minkowski or anti-de
Sitter spacetime) to zero, $\mathcal Q_\xi[g,\bar g]$ will provide
a well-defined quantity associated to $g$ and $\xi$. Let us now
explain how the construction works for some particular cases in
Einstein gravity.

\section{The energy in general relativity}

Let us only derive the conservation laws obtained originally by
Arnowitt, Deser and Misner \cite{Arnowitt:1962aa,Arnowitt:1962hi}
and Abbott and Deser \cite{Abbott:1981ff}. For simplicity, only
Einstein's gravity is discussed but other gauge theories admits
similar structures.

Let us first linearize the Einstein-Hilbert lagrangian $L^{EH}[g]$
with $g = \bar g +h$ around a solution $\bar g$. It can be shown
that the linearized lagrangian $L^{free}[h]$ is gauge invariant
under
\begin{equation}
\delta h_{\mu\nu} = \mathcal L_\xi \bar g_{\mu\nu},
\end{equation}
where $\xi$ is an arbitrary vector. Now, if the background $\bar
g_{\mu\nu}$ admits exact Killing vectors, the generalized Noether
theorem says that it exists $n-2$ forms $k_\xi[h,\bar g]$ which
are conserved when $h$ satisfies the linearized equations of
motion. In fact, the Killing vectors enumerate all non-trivial
solutions to $\mathcal L_\xi \bar g_{\mu\nu} \approx 0$ in that
case and the $n-2$ forms $k_\xi[h,\bar g]$ are thus the only
non-trivial conserved forms \cite{Barnich:2004ts}.

For Einstein gravity, the $n-2$ form associated to a Killing
vector $\xi$ of $\bar g$ is well-known to be
\cite{Abbott:1982ff,Iyer:1994ys,Anderson:1996sc,Wald:1999wa}
\begin{eqnarray}
  k_{\xi}[h,\bar g]=- \delta_h K_\xi[g] -
  i_\xi \Theta[h,\bar g] \label{GEO:eq:n-2forms}
  \end{eqnarray}
where $i_\xi = \xi^\mu \frac{\partial}{\partial dx^\mu}$ is the
inner product and the Komar term and the $\Theta$ term are given
by
\begin{eqnarray}
K_\xi[g] = \frac{\sqrt{-g}}{16\pi G} (D^\mu \xi^\nu - D^\nu
\xi^\mu )(d^{n-2}x)_{\mu\nu},\\
\Theta[h, g] = \frac{\sqrt{-g}}{16\pi G} (g^{\mu\alpha} D^\beta
h_{\alpha\beta} - g^{\alpha\beta} D^\mu h_{\alpha\beta})
(d^{n-1}x)_\mu.
\end{eqnarray}
Here, the variation $\delta_h$ acts on $g$ and the result of the
variation is evaluated on $\bar g$.

Now, the point is that this result may be lifted to the full
interacting theory (at least) in two different ways :
\begin{enumerate}
\item For suitable classes of spacetimes $(\mathcal M,g)$ with
\emph{boundary conditions} in an asymptotic region such that
\emph{the linearized theory applies} around some symmetric
background $\bar g$ in the asymptotic region.
\item For classes of solutions $(\mathcal M,g)$ with a set of exact Killing
vectors.
\end{enumerate}

To illustrate the first case, let take as an example
asymptotically flat spacetimes at spatial infinity $r \rightarrow
\infty$. This class of spacetimes is constrained by the condition
$g_{\mu\nu} - \eta_{\mu\nu} = O(1/r)$, where $\eta$ is the
Minkowski metric. We then consider the linearized field
$h_{\mu\nu} = g_{\mu\nu} - \eta_{\mu\nu}$ around the background
Minkowski metric. Under some appropriate additional boundary
conditions, it can be shown that the linearized theory applies at
infinity: since Minkowski spacetime admits Killing vectors $\bar
\xi$, $k_{\bar \xi}[h,\bar g]$\footnote{Note for completeness that
the theory of asymptotically conserved $n-2$ forms also allows for
an extended notion of symmetry, asymptotic symmetries, where
$\mathcal L_\xi \bar g$ tends to zero only asymptotically.} are
$n-2$-form conserved in the asymptotic region, i.e.
\emph{asymptotically conserved}\footnote{In fact, boundary
conditions are chosen such that the charges are finite, conserved
and form a representation of the Poincar\'e algebra.}. The
translations, rotations and boosts of Minkowski spacetime are thus
associated to energy-momentum and angular momentum. These are the
familiar ADM expressions. The conserved quantities in anti-de
Sitter spacetime can also be constructed that way.

In the second case, one applies the linearized theory around a
family of solution $g_{\mu\nu}$ which have $\xi$ as an exact
Killing vector. This allows one to compute the charge difference
between $g_{\mu\nu}$ and an infinitely close metric
$g_{\mu\nu}+\delta g_{\mu\nu}$. As $d k_\xi [\delta g,g]=0$ in the
whole spacetime, the charge difference $\oint_S k_\xi[\delta g,g]$
does not depend one the choice of integration surface,
\begin{equation}
\oint_S k_\xi[\delta g,g] = \oint_{S^\prime} k_\xi[\delta
g,g],\label{GEO:equal}
\end{equation}
where $S, S^\prime$ are any $n-2$ surfaces, usually chosen to be
$t=const$, $r=const$ in spherical coordinates. The total charge
associated to $\xi$ of a solution can then be defined by
\begin{equation}
Q_{\xi}[g,\bar g] =  \oint_S \int_{\bar g}^{g} k_\xi[\delta
g^\prime;g^\prime],
\end{equation}
where $\bar g_{\mu\nu}$ is a background solution with charge
normalized to zero and $g^\prime$ is the integration variable. The
outer integral is performed along a path of solutions. This
definition is only meaningful if the charge does not depend on the
path, which amount to what is called the \emph{integrability
condition}
\begin{equation}
\oint_S (\delta_1 k_\xi[\delta_2 g;g] - \delta_2 k_\xi[\delta_1
g;g] ) = 0.
\end{equation}

As a conclusion, we have sketched how one obtains the promised
definition of charge~\eqref{charge_gen} in the two aforementioned
cases. In the first (asymptotic) case, $K_\xi[g,\bar g] =
k_\xi[h,\bar g]$ where $h = g - \bar g$ is the linearized field at
infinity. In the second (exact) case, $K_\xi[g,\bar g] =
\int_{\bar g}^{g} k_\xi[\delta g^\prime;g^\prime]$, where one
integrates the linearized form $k_\xi$ along a path of solutions.

Finally note that all results presented here in covariant language
have their analogue in Hamiltonian form
\cite{Regge:1974zd,Anderson:1996sc,Barnich:2001jy}.

\chapter{Quasi-equilibrium states}

In 3+1 dimensions, stationary axisymmetric black holes are
entirely characterized by their mass and their angular momentum.
This fact is part of the \emph{uniqueness theorems}, see
\cite{Heusler:1998ua} for a review. In $n$ dimensions, the
situation is more complicated. First, the black hole may rotate in
different perpendicular planes. In 3+1 dimensions, the rotation
group $SO(3)$ has only one Casimir invariant, but in $n$
dimensions, it has $D \equiv \lfloor (n-1)/2 \rfloor$ Casimirs.
Therefore, one expects that, in general, a black hole will have
$D$ conserved angular momenta. This is what happens in the higher
dimensional Kerr and Reissner-Nordstr{\o}m black holes
\cite{Myers:1986un}. Remark that the generalization of rotating
black holes to anti-de Sitter backgrounds was done only very
recently \cite{Gibbons:2004uw}. So far so good; this is not a big
deal with respect to uniqueness.

The worrying (but interesting) point is that higher dimensions
allow for more exotic horizon topologies than the sphere. For
example, \emph{black ring} solutions where found
\cite{Emparan:2001wn} recently in 5 dimensions with horizon
topology $S^1 \times S^2$. The initial idea of the uniqueness
theorems that stationary axisymmetric black holes are entirely
characterized by a few number of charges at infinity is thus not
valid in higher dimensions. In what follows, we shall derive the
first law of black hole mechanics without using uniqueness
results.

From now, we restrict ourselves to stationary and axisymmetric
black holes with Killing horizon, having $\partial_t$ and
$\partial_{\varphi^a}$ , $1 \leq a \leq D$ as Killing vectors. We
allow for arbitrary horizon topology, only assuming the horizon is
connected. The Killing generator of the horizon is then a
combination of the Killing vectors,
\begin{equation}
\xi  = \frac{\partial}{\partial t} + \Omega^a
\frac{\partial}{\partial \varphi^a},
\end{equation}
where $\Omega^a$ are called the angular velocities at the horizon.

\section{The first law for Einstein gravity}

We are now set up to present the terms of the first law:
\begin{description} \item {\bf Theorem 5} \emph{First law.}
Let $(\mathcal M,g)$ and $(\mathcal M+\delta \mathcal M,g+\delta
g)$ be two slightly different stationary black hole solutions of
Einstein's equations with Killing horizon. The difference of
energy $\mathcal{E}$, angular momenta $\mathcal{J}_a$ and area
$\mathcal A$ of the black hole are related by
\begin{equation}
\delta \mathcal{E} = \Omega^a \, \delta \mathcal{J}_a +
\frac{\kappa}{8\pi} \delta \mathcal A,\label{GEO:first09}
\end{equation}
where $\Omega^a$ are the angular velocities at the horizon and
$\kappa$ is the surface gravity.
\end{description}
This \emph{equilibrium state version} of the first law of black
hole mechanics is essentially a balance sheet of energy between
two stationary black holes\footnote{It also exists a physical
process version, where an infinitesimal amount of matter is send
through the horizon from infinity.}. We shall prove that it comes
directly from the equality of the charge related to $\xi$ at the
horizon spacelike section $H$ and at infinity,
as~\eqref{GEO:equal},
\begin{eqnarray}
  \label{GEO:eq_first}
  \oint_{S^\infty} k_\xi[\delta
  g_{\mu\nu};g_{\mu\nu}]=\oint_{H} k_\xi[\delta g_{\mu\nu};g_{\mu\nu}].
\end{eqnarray}
The energy and angular momenta of the black hole are defined
as\footnote{The relative sign difference between the definitions
of $\mathcal E$ and $\mathcal J^a$ trace its origin to the Lorentz
signature of the metric \cite{Iyer:1994ys}.}
\begin{equation}
\delta \mathcal E = \oint_{S^\infty} k_{\partial_t}[\delta
  g_{\mu\nu};g_{\mu\nu}],\qquad \delta
\mathcal J^a= -\oint_{S^\infty} k_{\partial_{\varphi^a}}[\delta
  g_{\mu\nu};g_{\mu\nu}]\label{GEO:def_EJ}
\end{equation}
Therefore, the left-hand side of \eqref{GEO:eq_first} is by
definition given by
\begin{eqnarray}
   \oint_{S^\infty} k_\xi[\delta
  g_{\mu\nu};g_{\mu\nu}]=\delta \mathcal{E} - \Omega \delta
  \mathcal{J}\label{GEO:leftside}
\end{eqnarray}
Using~\eqref{GEO:eq:n-2forms}, we may rewrite the right-hand side
of \eqref{GEO:eq_first} as
\begin{eqnarray}
\oint_{H} k_\xi[\delta g_{\mu\nu};g_{\mu\nu}] = - \delta \oint_H
K_{\xi}[g] + \oint_H K_{\delta \xi } [g]-\oint_H \xi \cdot
\Theta[\delta g;g],\label{GEO:formula_charge}
\end{eqnarray}
where the variation of $\xi$ that cancels between the two first
terms on the right-hand side is put for later convenience.

On the horizon, the integration measure for $n-2$-forms is given
by
\begin{equation}
\sqrt{-g}(d^{n-2}x)_{\mu\nu} = \frac{1}{2} (\xi_\mu n_\nu - n_\mu
\xi_\nu) d\mathcal{A},\label{GEO:surf_form}
\end{equation}
where $d\mathcal{A}$ is the angular measure on $H$. Using the
properties of Killing horizons, the Komar integral on the horizon
becomes
\begin{equation}
\oint_H K_{ \xi}[g] = -\frac{\kappa \mathcal{A}}{8\pi G},
\end{equation}
where $\mathcal{A}$ is the area of the horizon. Now, it turns out
that
\begin{eqnarray}
\oint_H K_{\delta \xi }[g]-\oint_H \xi\cdot\Theta[\delta g;g]& =&
-\delta \kappa \frac{\mathcal{A}}{8\pi G}.\label{GEO:to_prove}
\end{eqnarray}
The computation which is straightforward but lengthly is done
explicitly in Appendix~\ref{GEO:app_firstlaw}\footnote{I thanks G.
Barnich for his suggestion of this computation.} \emph{without}
assuming specific invariance properties under the variation as
done in the original derivation~\cite{Bardeen:1973gs} and
subsequent derivations thereof \cite{Carter1973,Iyer:1994ys}.

The right-hand side of \eqref{GEO:eq_first} is finally given by
\begin{eqnarray}
\oint_{H} k_\xi[\delta
  g_{\mu\nu};g_{\mu\nu}] = \frac{\kappa}{8\pi G} \delta \mathcal{A},
\end{eqnarray}
as it should and the first law is proven. We can see in this
derivation that the first law is a \emph{geometrical} law in the
sense that it relates the geometry of Killing horizons to the
geometric measure of energy and angular momenta.

Remark finally that the derivation was done in arbitrary
dimensions, without hypotheses on the topology of the horizon and
for arbitrary stationary variations. The first law also applies in
particular for extremal black holes by taking $\kappa = 0$.

\section{Extension to electromagnetism}

It is straightforward to extend the present considerations to the
coupled Einstein-Maxwell system. The original derivation was given
in \cite{Bardeen:1973gs}, see also
\cite{Townsend:1997ku,Gao:2003ys} for alternative derivations. We
assume that no magnetic monopole is present, so that the potential
$A$ is regular everywhere.

According to the generalized Noether theorem, we first need to
extend the notion of symmetry. The gauge transformations of the
fields $(g_{\mu\nu},A_\mu)$ that are zero when the equations of
motion are satisfied are given by
\begin{eqnarray}
\mathcal{L}_\xi g_{\mu\nu} &=& 0,\\
\mathcal{L}_\xi A_\mu + \partial_\mu \epsilon &=&
0.\label{GEO:em_symm}
\end{eqnarray}
These equations are the generalized Killing equations. We consider
only stationary and axisymmetric black holes with Killing horizon
which have as a solution to these equations $(\xi,\epsilon) =
(\partial_t,0)$, $(\xi,\epsilon) = (\partial_{\varphi^a},0)$ for
$a = 1...D$ and $(\xi,\epsilon) = (0,\epsilon \in \mathbb R)$. The
conserved quantities associated to these symmetry parameters are
respectively the energy, the angular momenta and the electric
charge. The existence of the electric charge independently from
the other charges suggests that it will show up in the first law.

The conserved superpotential associated to a symmetry parameter
$(\xi,\epsilon)$ can be shown to be
\begin{equation}
k^{tot}_{\xi,\epsilon}[\delta g,\delta A; g,A]= k^{grav}_{\xi}
-\delta K^{em}_{\xi,\epsilon } + K^{em}_{\delta\xi,\delta\epsilon
} - \xi \cdot \Theta^{em}\label{GEO:kem}
\end{equation}
where $k^{grav}_{\xi}[\delta g,g]$ is the gravitational
contribution~\eqref{GEO:eq:n-2forms}\footnote{We use ``geometrized
units'' for the electromagnetic field; the lagrangian is $L =
\frac{\sqrt{-g}}{16\pi G }(R - \frac{1}{4}F^2)$.},
\begin{equation}
K^{em}_{\xi,\epsilon}[g,A]=\frac{\sqrt{-g}}{16\pi G}
[F^{\mu\nu}(\xi^\alpha A_\alpha+\epsilon)](d^{n-2}x)_{\mu\nu}
\end{equation}
and
\begin{equation}
\Theta^{em}[g,A ; \delta A]=\frac{\sqrt{-g}}{16\pi G}
F^{\alpha\mu}\delta A_\alpha (d^{n-1}x)_{\mu}.
\end{equation}
Let us now look at the fundamental equality~\eqref{GEO:eq_first}
of the first law where we choose $\xi$ as the Killing generator
and $\epsilon = 0$. The superpotential $k = k^{tot}$ should
contain the electromagnetic contributions as well.

On the one hand, the energy and angular momenta are still defined
by \eqref{GEO:def_EJ} where $k$ is given by \eqref{GEO:kem}. For
usual potentials, the electromagnetic contributions vanish at
infinity and only the gravitational contributions are
important\footnote{Note however that if the gauge potential tends
to a constant at infinity, the quantities~\eqref{GEO:def_EJ} will
contain a contribution from the electric charge. We choose for
convenience a potential vanishing at infinity. The contributions
from the electromagnetic fields will then comes only from the
surface integral over the horizon.}. Equation~\eqref{GEO:leftside}
still hold. On the other hand, at the horizon, the electromagnetic
field is not negligible and we have
\begin{equation}
\oint_{H} k^{tot}_{\xi,0} = \frac{\kappa}{8\pi} \delta \mathcal{A}
-\delta\oint_{H} K^{em}_{\xi,\epsilon } +
\oint_{H}K^{em}_{\delta\xi,0 } - \oint_{H} \xi \cdot
\Theta^{em}.\label{GEO:first_EM}
\end{equation}
Now, remember that the zero law extended to electromagnetism said
that $\Phi = -(\xi^\alpha A_\alpha)|_H$ is constant on the
horizon\footnote{We previously assumed that $\mathcal L_\xi A_\mu
= 0$. So $\epsilon =0$ here.}. Therefore, we have directly
\begin{equation}
\oint_H K^{em}_{\xi,0} = -\Phi \mathcal{Q}.
\end{equation}
In order to work out the two remaining terms on the right hand
side of~\eqref{GEO:first_EM}, let us rewrite
equation~\eqref{GEO:em_symm} as
\begin{equation}
\mathcal{L}_\xi A_\mu = \xi^\nu F_{\nu \mu} + (\xi^\alpha
A_\alpha)_{,\mu} = 0.
\end{equation}
Since $\Phi$ is constant on the horizon, this equation says that
the electric field $E^\mu \equiv F^{\mu\nu}\xi_\nu$ with respect
to $\xi$ satisfy $\eta_\mu E^\mu \overset{\mathcal{H}}{=} 0$ for
all tangent vectors $\eta^\mu$, therefore $E^\mu$ is proportional
to $\xi^\mu$ or more precisely,
\begin{equation}
F^{\mu\nu}\xi_\nu \overset{\mathcal{H}}{=} (F^{\nu\rho}\xi_\nu
n_\rho)\xi^\mu \label{GEO:elec}.
\end{equation}
Using the last relation with~\eqref{GEO:surf_form}, it is easy to
show that
\begin{eqnarray}
\oint_H K^{em}_{\delta\xi,0} - \oint_H \xi \cdot \Theta^{em} &=&
-\delta \Phi \mathcal{Q} - \oint_H \frac{d\mathcal{A}}{4\pi G}
n_\mu F^{\mu\nu}\delta A_\nu \xi^2.
\end{eqnarray}
The last term vanishes because of \eqref{GEO:xi0}. Finally, we
obtain the first law valid when electromagnetic fields are
present,
\begin{eqnarray}
  \label{GEO:eq:15bis}
  \delta\mathcal{E} -\Omega_a\delta \mathcal{J}^a= \frac{\kappa}{8\pi G}
   \delta \mathcal{A}+\Phi \delta \mathcal{Q}.
\end{eqnarray}

Note finally that the first applies for any theory of gravity,
from arbitrary diffeomorphic invariant Lagrangians
\cite{Iyer:1994ys}, to black holes in non-conventional background
geometries \cite{Gibbons:2004ai,Barnich:2005kq} or to black
objects in string theory or supergravity
\cite{Horowitz:1993jc,Copsey:2005se}.

In addition, Hawking \cite{Hawking:1974sw} discovered that $T
\equiv \frac{\kappa \hbar}{2\pi}$ is the temperature of the
quantum radiation emitted by the black hole. Moreover, as tells us
the second law, $S \equiv \frac{\mathcal A}{4 \hbar G}$ is a
quantity that can classically only increase with time. It suggests
\cite{Bekenstein:1974ax} to associate an entropy $S$ to the black
hole. This striking occurrence of thermodynamics emerging out of
the structure of gravity theories is a fundamental issue to be
further explained in still elusive quantum gravity theories.

\section{Extension to any theory of gravity}

Let us finally briefly review the proposal of Iyer and Wald
\cite{Iyer:1994ys,Iyer:1995kg} for the extension of the first law
to an arbitrary diffeomorphic invariant Lagrangian. Let
\begin{equation}
L[\phi] = L[g_{\mu\nu},R_{\mu\nu\rho\sigma;(\alpha)}],
\end{equation}
be the lagrangian where $\phi$ denote all the fields of the theory
and $(\alpha)$ denotes an arbitrary number of symmetrized
derivatives.

According to the authors, the charge difference between the
solutions $\phi$ and $\phi+\delta \phi$ corresponding to the
vector $\xi$ can be written as \eqref{GEO:formula_charge}
\cite{Iyer:1994ys,Iyer:1995kg} where the Komar term $\oint K_\xi$
is given by
\begin{eqnarray}
\oint (d^{n-2}x)_{\mu\nu}\left(\sqrt{-g}\xi^\mu
W^\nu[\phi]+Y^{\mu\nu}[\mathcal L_\xi \phi,\phi] - \frac{\delta L
}{\delta R_{\mu\nu \alpha\beta}}\xi_{\alpha;\beta} - (\mu
\leftrightarrow \nu)\hspace{-3pt}\right) \label{GEO:Komargen}
\end{eqnarray}
and where $\Theta[\delta \phi ;\phi]$, $W^\mu[\phi]$ and
$Y^{\mu\nu}[\mathcal L_\xi \phi,\phi]$ have a generic form which
we shall not need here. The energy and angular momentum are
defined as previously by~\eqref{GEO:def_EJ} where the $n-2$ form
$k_\xi$ is defined by \eqref{GEO:formula_charge}.

Therefore,
equations~\eqref{GEO:eq_first}-\eqref{GEO:leftside}-\eqref{GEO:formula_charge}
are still valid with appropriate $K_\xi[\phi]$ and $\Theta[\delta
\phi ;\phi]$. Now, we assume that the surface gravity $\kappa$ is
not vanishing and that the horizon generators are geodesically
complete to the past. Then, it exists \cite{Racz:1992bp} a special
$n-2$ spacelike surface on the horizon, the bifurcation surface,
where the Killing vector $\xi$ vanish. For Killing vectors $\xi$,
i.e. such that $\mathcal L_\xi \phi^i = 0$ and for solutions of
the equations of motion, $k_\xi$ is a closed
form,~\eqref{GEO:equal} hold and we may evaluate the right-hand
side of~\eqref{GEO:formula_charge} on the bifurcation surface.
There, assuming regularity conditions, only the third term in the
Komar expression~\eqref{GEO:Komargen} contributes
\cite{Iyer:1994ys,Jacobson:1993vj} and the right-hand side
of~\eqref{GEO:formula_charge} becomes
\begin{equation}
\frac{\kappa}{2\pi}\delta \mathcal S,
\end{equation}
where the ``higher order in the curvature'' entropy is defined by
\begin{equation}
\mathcal S = - 8\pi \oint_H (d^{n-2}x)_{\mu\nu} \frac{\delta L
}{\delta R_{\mu\nu \alpha\beta}}\xi_\alpha n_\beta
\label{GEO:highestS}.
\end{equation}
The first law~\eqref{GEO:first09} therefore holds with appropriate
definitions of energy, angular momentum and entropy.

For the Einstein-Hilbert lagrangian, we have
\begin{equation}
\frac{\delta L^{EH} }{\delta R_{\mu\nu \alpha\beta}} =
\frac{\sqrt{-g}}{32\pi G} (g^{\mu\alpha}g^{\nu
\beta}-g^{\mu\beta}g^{\nu\alpha}),
\end{equation}
and the entropy~\eqref{GEO:highestS} reduces to the familiar
expression $\mathcal A /4G$.

\section*{Acknowledgments} I am grateful to G. Barnich, V. Wens,
 M. Leston and S. Detournay for their reading of the
manuscript and for the resulting fruitful discussions. Thanks also
to all participants and organizers of the Modave school for this
nice time. This work is supported in part by a ``P{\^o}le
d'Attraction Interuniversitaire'' (Belgium), by IISN-Belgium,
convention 4.4505.86, by the National Fund for Scientific Research
(FNRS Belgium), by Proyectos FONDECYT 1970151 and 7960001 (Chile)
and by the European Commission program MRTN-CT-2004-005104, in
which the author is associated to V.U.~Brussel.

\appendix

\chapter{Completion of the proof of the first law}
\label{GEO:app_firstlaw}

Let us prove the relation~\eqref{GEO:to_prove}. We consider any
stationary variation of the fields $\delta g_{\mu\nu}$, $\delta
\xi^\mu$, i.e. such that
\begin{equation}
\mathcal L_\xi \delta g_{\mu\nu} + \mathcal L_{\delta \xi}
g_{\mu\nu} = 0.\label{GEO:varKill}
\end{equation}
The variation is chosen to commute with the total derivative, i.e.
the coordinates are left unchanged $\delta x^\mu = 0$.

Using the decomposition~\eqref{GEO:surf_form}, the left-hand side
of equation~\eqref{GEO:to_prove} can written explicitly as
\begin{eqnarray}
\oint_H K_{\delta \xi }[g]-\oint_H \xi\cdot\Theta[\delta g;g]& =&
\oint_{H} \frac{d\mathcal{A}}{16\pi G} \Big(
-\delta\xi^{\mu;\nu}(\xi_\mu n_\nu - \xi_\nu n_\mu) \nonumber\\
&&+ \xi^\mu (\delta
g_{\mu\nu}^{\,\,\,\,\,\,\,;\nu}-g^{\alpha\beta}\delta
g_{\alpha\beta;\mu}) \Big).\label{GEO:topr}
\end{eqnarray}
We have to relate this expression to the variation of the surface
gravity $\kappa$. This is merely an exercise of differential
geometry.

Since the horizon $S(x) = 0$ stay at the same location in $x^\mu$,
the covariant vector normal to the horizon $\xi_\mu = f
\partial_\mu S$, where $f$ is a $\kappa$-dependent normalization
function, satisfies
\begin{equation}
\delta \xi_\mu \overset{\mathcal{H}}{=} \delta f
\xi_\mu,\label{GEO:dxi}
\end{equation}
where $\delta \xi_\mu \equiv \delta (g_{\mu\nu}\xi^\nu)$. From the
variation of \eqref{GEO:xi0} and of the second normalization
condition~\eqref{GEO:norm}, one obtains
\begin{equation}
\delta \xi^\mu \xi_\mu \overset{\mathcal H}{=} 0, \qquad \delta
n^\mu \xi_\mu \overset{\mathcal H}{=} \delta
f,\label{GEO:rel_delta}
\end{equation}
which shows that $\delta \xi^\mu$ has no component along $n^\mu$
and $\delta n^\mu$ has a component along $n^\mu$ which equals
$-\delta f$\footnote{Note also the following property that is
useful in order to prove the first law in the way of
\cite{Iyer:1994ys}. Using~\eqref{GEO:defq} and \eqref{GEO:dxi}, we
have
\begin{equation}
\delta \xi_{\mu,\nu} - \delta \xi_{\nu,\mu} \overset{\mathcal
H}{=} \xi_\mu (\delta f q_\nu + \delta q_\nu) - \xi_\nu (\delta f
q_\mu + \delta q_\mu).
\end{equation}
It implies in particular that the expression $\delta
\xi_{[\mu;\nu]}$ has no tangential-tangential component, $\delta
\xi_{[\mu;\nu]}\eta^\mu \tilde \eta^\nu \overset{\mathcal H}{=}
0$, $\forall \eta, \tilde \eta$ orthogonal to $\mathcal H$.}.

Let us develop the variation of $\kappa$ starting from the
definition~\eqref{GEO:kappa}. One has
\begin{eqnarray}
\delta \kappa &=& \frac{1}{2}(\xi^\mu \xi_\mu)_{;\nu} \delta n^\nu
+ \frac{1}{2} (\delta \xi_\mu \xi^\mu + \xi_\mu \delta \xi^\mu
)_{;\nu}n^\nu, \\
&=& \frac{1}{2} \delta \xi_{\mu ; \nu}(\xi^\mu n^\nu + \xi^\nu
n^\mu) + \xi^\mu_{\,\,\, ;\nu}(\delta \xi_\mu n^\nu + \xi_\mu
\delta
n^\nu)\nonumber \\
&& + \frac{1}{2}n^\nu  (\xi_\mu \delta \xi^\mu)_{;\nu}-\frac{1}{2}
n^\nu \mathcal{L}_\xi \delta \xi_\nu, \label{GEO:eq_kappa}
\end{eqnarray}
where all expressions are implicitly pulled-back on the horizon.
The first term in~\eqref{GEO:eq_kappa} is recognized as
$-\frac{1}{2} \delta \xi_{\mu ; \nu}g^{\mu\nu}$ after
using~\eqref{GEO:metric}, \eqref{GEO:dxi} and \eqref{GEO:ortho}.
According to \eqref{GEO:dxi}-\eqref{GEO:rel_delta}, the second
term can be written as
\begin{equation}
\xi^\mu_{\,\,\, ;\nu}(\delta \xi_\mu n^\nu + \xi_\mu \delta n^\nu)
= \xi_{\mu ; \nu} \xi^\mu \delta\eta^\nu,
\end{equation}
for some $\delta \eta^\nu$ tangent to $\mathcal{H}$. This term
vanishes thanks to~\eqref{GEO:ortho}. The third term can be
written as
\begin{eqnarray}
\frac{1}{2}n^\nu (\xi_\mu \delta \xi^\mu)_{;\nu} = -\frac{1}{2}
n^\nu \mathcal{L}_{\delta \xi}\xi_\nu + n_\nu \xi_\mu \delta
\xi^{\mu ; \nu}.
\end{eqnarray}
Now, the Lie derivative of $\delta \xi_\mu$ along $\xi$ can be
expressed as
\begin{equation}
\mathcal{L}_\xi \delta \xi_\mu = -\mathcal{L}_{\delta
\xi}\xi_\mu,\label{GEO:Lie_diff}
\end{equation}
by using the Killing equation~\eqref{GEO:Kill} and its
variation~\eqref{GEO:varKill}. The fourth term can then be written
as
\begin{equation}
-\frac{1}{2} n^\nu \mathcal{L}_\xi \delta \xi_\nu= \frac{1}{2}
n^\nu \mathcal{L}_{\delta \xi} \xi_\nu.
\end{equation}
Adding all the terms, the variation of the surface gravity becomes
\begin{eqnarray}
\delta \kappa &=& -\frac{1}{2} (\delta \xi_\mu)^{;\mu} + \delta
\xi^{\mu;\nu}\xi_\mu n_\nu ,\nonumber\\
&=& -\frac{1}{2} \delta g_{\mu\nu}^{\,\,\,\,\,\,\,;\mu}\xi^\nu -
\frac{1}{2} \delta \xi^\mu_{\,\,\, ;\mu} + \frac{1}{2}
\delta\xi^{\mu;\nu}(\xi_\mu n_\nu - \xi_\nu n_\mu) + \frac{1}{2}
\delta\xi^{\mu;\nu}(\xi_\mu n_\nu + \xi_\nu
n_\mu),\nonumber\\
&=& -\frac{1}{2} \delta g_{\mu\nu}^{\,\,\,\,\,\,\,;\mu}\xi^\nu -
\delta \xi^\mu_{\,\,\, ;\mu} + \frac{1}{2}
\delta\xi^{\mu;\nu}(\xi_\mu n_\nu - \xi_\nu n_\mu) + \frac{1}{2}
\delta \xi^{\mu;\nu}\gamma_{\mu\nu}.\label{GEO:lastt}
\end{eqnarray}
The last line is a consequence of \eqref{GEO:metric}. Contracting
\eqref{GEO:varKill} with $g^{\mu\nu}$ we also have
\begin{equation}
\delta \xi^\mu_{\;\,;\mu} = -\frac{1}{2} \xi^\mu g^{\alpha
\beta}\delta g_{\alpha\beta ;\mu}.
\end{equation}
Finally, the last term in \eqref{GEO:lastt} reduces to
$\frac{1}{2} \delta t^{\alpha}_{\,\,\, |\alpha} $ where $|\alpha$
denotes the covariant derivative with respect to the $n-2$ metric
$\gamma_{\mu\nu}$ and $\delta t^{\mu} = \gamma^\mu_{\,\,\, \nu
}\delta \xi^\nu$ is the pull-back of $\delta \xi^\mu$ on
$\mathcal{H}$. Indeed, one has
\begin{eqnarray}
\frac{1}{2} \delta \xi^{\mu;\nu}\gamma_{\mu\nu} &=& \frac{1}{2}
\delta
t^{\mu;\nu}\gamma_{\mu\nu},\\
&=& \frac{1}{2} (\delta t^{\mu}_{\;\, ,\nu}\gamma_\mu^{\;\,\nu} +
\Gamma_{\mu; \nu
\alpha}\gamma^{\mu\nu}\delta t^\alpha),\\
&=& \frac{1}{2} \delta t^\mu_{\,\,\, |\mu},
\end{eqnarray}
where $|\mu$ denotes the covariant derivative with respect to the
$n-2$ metric $\gamma_{\mu\nu}$. The first line
uses~\eqref{GEO:Kill}-\eqref{GEO:rel_delta} and
$\gamma_{\mu\nu}\xi^\nu=0=\gamma_{\mu\nu}n^\nu$. The last line
uses the decomposition~\eqref{GEO:metric} and $\delta t^\alpha
n_\alpha= 0 =\delta t^\alpha \xi_\alpha$. We have finally the
result
\begin{equation}
\delta \kappa =  -\frac{1}{2} \delta
g_{\mu\nu}^{\,\,\,\,\,\,\,;\mu}\xi^\nu + \frac{1}{2} \xi^\mu
g^{\alpha \beta}\delta g_{\alpha\beta ;\mu} + \frac{1}{2}
\delta\xi^{\mu;\nu}(\xi_\mu n_\nu - \xi_\nu n_\mu) + \frac{1}{2}
(\gamma^\mu_{\,\,\,
\nu}\delta\xi^\nu)_{|\mu}.\label{GEO:deltakappa}
\end{equation}
Expression~\eqref{GEO:topr} is therefore equals to
\begin{eqnarray}
\oint_H K_{\delta \xi }[g]-\oint_H \xi\cdot\Theta[\delta g;g]& =&
-\oint_{H} \frac{d\mathcal{A}}{8\pi G} \delta \kappa,
\end{eqnarray}
and the result~\eqref{GEO:to_prove} follows because $\delta
\kappa$ is constant on the horizon.

Remark that in classical derivations
\cite{Bardeen:1973gs,Carter1973}, it is assumed that the Killing
vectors $\partial_t$ and $\partial_\varphi$ have the same
components before and after the variation,
\begin{equation*}
\delta (\partial_t)^\mu = \delta (\partial_\varphi)^\mu = 0.
\end{equation*}
One then has $\delta \xi^\mu = \delta \Omega
(\partial_\varphi)^\mu$ and the variation of $\kappa$ reduces to
the well-known expression
\begin{equation*}
\delta \kappa =  -\frac{1}{2} \delta
g_{\mu\nu}^{\,\,\,\,\,\,\,;\mu}\xi^\nu + \delta \Omega
(\partial_\varphi)^{\mu;\nu}\xi_\mu n_\nu.
\end{equation*}

\providecommand{\href}[2]{#2}\begingroup\raggedright\endgroup

\end{document}